\documentclass[a4paper]{article}

\usepackage{INTERSPEECH2019}

\usepackage{amsmath,graphicx} %,graphicx}
\usepackage{amssymb,amsfonts}

\usepackage{color}
\usepackage[pagebackref, colorlinks, linkcolor=red, citecolor=green]{hyperref}
\usepackage{url}
\usepackage{cite}
\usepackage{setspace}

\newcommand{\vm}[1]{\textit{\boldmath $#1$}}
\newcommand{\x}{\vm{x}}
\newcommand{\y}{\vm{y}}
\newcommand{\pdf}{p.d.f.}

\title{WaveCycleGAN2: Time-domain Neural Post-filter \\ for Speech Waveform Generation}
\name{Kou Tanaka, Hirokazu Kameoka, Takuhiro Kaneko, Nobukatsu Hojo}
\address{NTT Communication Science Laboratories, NTT Corporation, Japan}
\email{\{kou.tanaka.ef, hirokazu.kameoka.uh, \\takuhiro.kaneko.tb, nobukatsu.houjou.ae\}@hco.ntt.co.jp}

\begin{document}

\maketitle
\begin{abstract}
WaveCycleGAN has recently been proposed to bridge the gap between natural and synthesized speech waveforms in statistical parametric speech synthesis and provides fast inference with a moving average model rather than an autoregressive model and high-quality speech synthesis with the adversarial training.
However, the human ear can still distinguish the processed speech waveforms from natural ones.
One possible cause of this distinguishability is the aliasing observed in the processed speech waveform via down/up-sampling modules. 
To solve the aliasing and provide higher quality speech synthesis, we propose WaveCycleGAN2, which 1) uses generators without down/up-sampling modules and 2) combines discriminators of the waveform domain and acoustic parameter domain.
The results show that the proposed method~\footnote{\label{foot1}Audio samples can be accessed on our web page: \href{http://www.kecl.ntt.co.jp/people/tanaka.ko/projects/wavecyclegan2/index.html}{www.kecl.ntt.co.jp/people/tanaka.ko/projects/wavecyclegan2/index.html}}
1) alleviates the aliasing well,
2) is useful for both speech waveforms generated by analysis-and-synthesis and statistical parametric speech synthesis,
and 3) achieves a mean opinion score comparable to those of natural speech and speech synthesized by WaveNet~\cite{oord2016wavenet} (open WaveNet~\cite{ryuichiURLwavenet}) and WaveGlow~\cite{prenger2018waveglow} while processing speech samples at a rate of more than 150 kHz on an NVIDIA Tesla P100.
\end{abstract}

\section{Introduction}
Speech processing systems using classical parametric vocoder frameworks such as STRAIGHT~\cite{kawahara1999restructuring, kawahara2001aperiodicity} and WORLD~\cite{morise2016world} are popular frameworks for various tasks such as statistical parametric speech synthesis (SPSS)~\cite{zen2013statistical} and statistical voice conversion (VC)~\cite{toda2007voice}.
The key advantage of using the classical parametric vocoder frameworks is that speech signals can be represented by interpretable and compact acoustic parameters such as the fundamental frequency ($F_0$) and Mel-cepstrum rather than a short-term Fourier transform (STFT) spectrogram.
On the other hand, the critical drawback is that the generated speech can usually be distinguished from natural speech due to vocoding error~\cite{zen2009statistical}, even through we only re-synthesize the speech waveform from the analyzed acoustic parameter.
Moreover, operating acoustic parameters and their statistics in SPSS and VC usually leads to an over-smoothing effect~\cite{zen2009statistical} and increases the differences between synthetic and natural speech.

To address these drawbacks, we previously proposed a learning-based filter of the time-domain, called WaveCycleGAN~\cite{tanaka2018wavecyclegan}, which allows us to convert a synthetic speech waveform into a natural speech waveform using cycle-consistent adversarial networks with a fully convolutional architecture.
The difficulties of the waveform conversion within the deep learning approaches are the difficulty of parallel data collection of speech waveform and the ambiguity of the phase information of speech waveform.
Notably, the phase ambiguity prevents proper learning of a mapping function from synthetic speech to natural speech (e.g., when we have two speech waveforms with certain phase spectra and the reversed phase spectra in the training data of natural speech, minimizing the objective function results in converting to silence).
The cyclic model makes it possible to address these difficulties of the operating waveform.
Moreover, WaveCycleGAN is trained within the adversarial learning, so no explicit assumption against speech waveform is required.
As a result, by applying WaveCycleGAN to SPSS trained for a Japanese dataset, the filtered speech has achieved a mean opinion score higher than 4.
However, there is still a gap between natural speech and filtered speech.
In the preliminary experiment, by applying WaveCycleGAN to the speech waveform synthesized from acoustic parameters of natural speech, the filtered speech was lower quality than the input of WaveCycleGAN.
We found that one possible reason for the difference and degradation in quality is the aliasing observed in the filtered speech waveform via down/up-sampling modules in model architectures.

To bridge the gap including the aliasing, we propose WaveCycleGAN2, which is an improved variant of WaveCycleGAN that 
1) uses generators without down/up-sampling modules 
and 2) combines discriminators of the waveform domain and acoustic parameter domain such as Mel-spectrogram, Mel-frequency cepstral coefficients, and phase spectrum. 
We analyzed the effect of each technique on an internal Japanese speech dataset~\footnote{\label{foot2} On our web page, we used an alternative speech dataset, the CMU Arctic database~\cite{kominek2004cmu}, which allows us to publish} and a public domain English speech dataset~\cite{ljspeech17}.
Experimental evaluations showed that the proposed method 
1) alleviates the aliasing well, 
2) is useful for both speech waveforms generated by analysis-and-synthesis (AnaSyn) and SPSS, 
and 3) achieves a mean opinion score comparable to those of natural speech and speech synthesized by WaveNet~\cite{oord2016wavenet} (open WaveNet~\cite{ryuichiURLwavenet}) and WaveGlow~\cite{prenger2018waveglow} while processing audio samples at a rate of more than 150 kHz on an NVIDIA Tesla P100.

\section{Related Works}
\subsection{Vocoder for Waveform Generation}
To generate a speech waveform from given acoustic parameters, neural-network-based waveform models~\cite{oord2016wavenet,kalchbrenner2018efficient,prenger2018waveglow,wang2018neural} have been proposed and have performed outstandingly at numerous tasks involving speech signal processing. 
There are two types of neural-network-based waveform models: an autoregressive (AR) model~\cite{oord2016wavenet,kalchbrenner2018efficient} and a moving-average (MA) model~\cite{prenger2018waveglow,wang2018neural} (a.k.a., non-AR model). 
As an AR model, although the WaveNet~\cite{oord2016wavenet} synthesizes speech with high fidelity, its training procedure is complex~\footnote{\label{foot3} Generated speech sometimes collapses as reported by Wu et al. ~\cite{wu2018collapsed}.} and its inference speed is quite slow because the AR process never allows us to infer several waveform sampling points in parallel.

For MA models that allow us to parallelize the inference, distilled models~\cite{oord2017parallel,ping2018clarinet} demanding complex training criteria have also been proposed. 
To avoid the complex training criteria, WaveGlow~\cite{prenger2018waveglow} is a theoretically powerful model that has the tractability of the exact log-likelihood. 
Although WaveGlow makes it possible to train the theoretically exact mapping function by using only one criterion, it requires large-scale computational resources and long training time. 
To make the inference procedure interpretable, a neural source filter model~\cite{wang2018neural} has also been proposed as an MA model. 
All these models can work well if the given acoustic parameters are close to natural ones seen in the training data. 
Otherwise, the training procedure has to be several steps rather than one step because it is combined with other training procedures such as fine-tuning~\cite{tobing2019voice} or methods described in the next subsection. 
In contrast, even if the given acoustic parameters are not close to natural ones in the training data, our approach, which is a kind of the MA model, makes it possible to directly train the mapping function from the generated speech waveform to the natural one in one step because it allows the use of a classical parametric vocoder that is not necessary to train.

\subsection{Acoustic Parameter Generation/Modification}
To address the over-smoothing effect~\cite{zen2009statistical}, several techniques have been proposed~\cite{toda2007voice, takamichi2014postfilter, kaneko2017generative} to restore the fine structure of acoustic parameters of natural speech~\footnote{\label{foot4} Of course, accurate modeling approaches have also been proposed, such as generative adversarial network-based text-to-speech~\cite{saito2018statistical,ma2018a} and voice conversion~\cite{kameoka2018stargan}.}. 
In their respective directions, these approaches have significantly improved the naturalness of acoustic parameters generated by SPSS and VC. 
However, these approaches are still insufficient to generate natural-sounding speech because of the post-filter of acoustic parameters on the heuristically limited compact space even in the generative adversarial nets (GAN) based approach~\cite{kaneko2017generative}. 
Moreover, it is impossible to address the vocoding error~\cite{zen2009statistical} when we use the classical parametric vocoder to generate the speech waveform. 
In contrast, our approach allows us to address both the over-smoothing effect and vocoding error because of the processing after waveform generation processing.

\section{Conventional WaveCycleGAN}
We briefly review our previous work, WaveCycleGAN~\cite{tanaka2018wavecyclegan}, which is a kind of cyclic model (a.k.a., dual learning~\cite{he2016dual}).

Let us use one-dimensional vectors $\x$ and $\y$ to denote sequences belonging to sets of synthetic $X$ and natural $Y$ speech waveforms, respectively. 
Inspired by CycleGAN~\cite{zhu2017unpaired}, WaveCycleGAN uses three training criteria (adversarial loss ${\mathcal L}_\mathrm{adv}$~\cite{goodfellow2014generative}, cycle-consistency loss ${\mathcal L}_\mathrm{cyc}$~\cite{zhou2016learning}, and identity-mapping loss ${\mathcal L}_\mathrm{id}$~\cite{taigman2016unsupervised}) to train a mapping function $G_{X\to Y}$ that converts the waveform of synthetic speech into that of natural speech without relying on parallel data.

The adversarial loss is written as,
\begin{align}
  {\mathcal L}_\mathrm{adv} (G_{X\to Y}, & D_Y) = \mathbb{E}_{y \sim P_Y(y)} [\log D_Y (\y)] \nonumber \\
     & + \mathbb{E}_{x \sim P_X(x)} [\log (1 - D_Y(G_{X\to Y}(\x)))], \label{eq:loss_adv}
\end{align}
\noindent where $D_Y$ indicates a discriminator trying to differentiate between a real sample $\y$ and the samples $G_{X\to Y}(\x)$ converted by the generator $G_{X\to Y}$ while $G_{X\to Y}$ is trained for converting $\x$ to $G_{X\to Y}(\x)$ that can deceive $D_Y$ as $\y$. 
This criterion focuses on only whether it can deceive $D_Y$ or not, so $\x$ might be converted into samples that have different linguistic information. To retain the linguistic information of the input $\x$, the cycle-consistency and identity-mapping losses are used:
\begin{align}
    {\mathcal L}_\mathrm{cyc} & = \mathbb{E}_{x \sim P_X(x)} [|| G_{Y \to X}(G_{X \to Y}(\x)) - \x ||_1] \nonumber \\
    & + \mathbb{E}_{y \sim P_Y(y)} [|| G_{X \to Y}(G_{Y \to X}(\y)) - \y ||_1], \\
    {\mathcal L}_\mathrm{id} & = \mathbb{E}_{y \sim P_Y(y)} [|| G_{X \to Y}(\y) - \y ||_1] \nonumber \\
    & + \mathbb{E}_{x \sim P_X(x)} [|| G_{Y \to X}(\x)) - \x ||_1],
\end{align}
\noindent where $G_{Y \to X}$ indicates another generator that has the reverse direction to $G_{X \to Y}$. 
Note that to guide the learning direction, ${\mathcal L}_\mathrm{id}$ is usually used only in the early stage of the training.

Finally, the full objective function can be written as
\begin{align}
    {\mathcal L}_\mathrm{full} & = {\mathcal L}_\mathrm{adv}(G_{X\to Y}, D_Y) + {\mathcal L}_\mathrm{adv}(G_{Y\to X}, D_X) \nonumber \\ 
    & + \lambda_\mathrm{cyc}{\mathcal L}_\mathrm{cyc} + \lambda_\mathrm{id}{\mathcal L}_\mathrm{id},
\end{align}
\noindent where $\lambda_\mathrm{cyc}$ and $\lambda_\mathrm{id}$ indicate hyperparameters controlling the cycle-consistency and identity-mapping losses.

\section{Proposed WaveCycleGAN2}

\subsection{Aliasing Issue of Conventional WaveCycleGAN}
\label{subsec:aliasing}
\begin{figure*}[t]
    \centering
    \includegraphics[width=170mm]{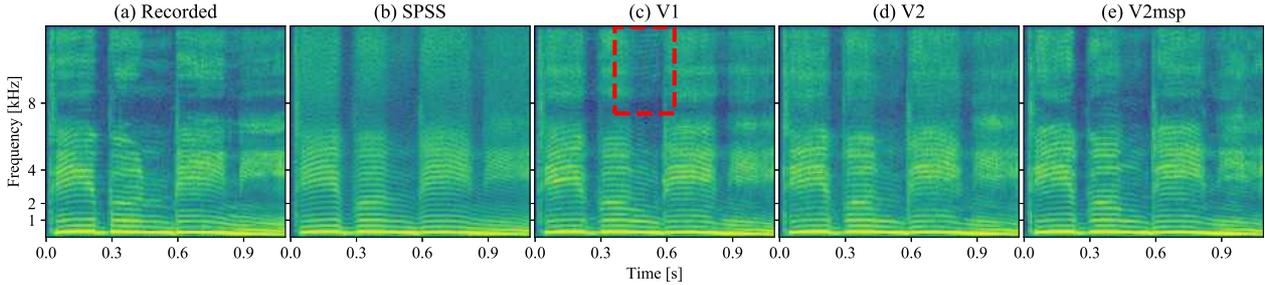}
    \caption{Spectrogram of natural waveform and waveforms generated from models in Sec.~\ref{subsubsec:systems}. Dashed red box indicates aliasing.}
    \label{fig:spectrum}
\end{figure*}

Many model architectures using convolutional neural networks involve down/up-sampling modules as the de facto standard~\cite{he2016deep,ledig2016photo,sun2018fishnet,ravanelli2018speaker} because it has a significant advantage in terms of the computation amount. 
We also adopted convolution with strides to WaveCycleGAN because of its computational advantage. 
As a result, we achieved a mean opinion score higher than 4 in terms of the naturalness. 
However, we found that aliasing is observed in the processed speech waveform, as shown in Fig.~\ref{fig:spectrum} (c). 
This phenomenon has also been reported in several other tasks such as image classification~\cite{zeiler2014visualizing} and deep speech processing~\cite{gong2018impact}. 
The aliasing occurs when the Nyquist-Shannon sampling theorem~\cite{shannon1998communication} is not satisfied, so it follows that the classical convolution with strides is not guaranteed to satisfy the sampling theorem. 
This is reasonable because the classical convolution with strides is not guaranteed to have an anti-aliasing mechanism while we never perform down-sampling without anti-aliasing processing in the pure signal processing. 
Note that in acoustic parameter trajectory smoothing~\cite{takamichi2015naist}, the acoustic parameter differences in high modulation frequency~\footnote{\label{foot5} Modulation frequency is the frequency of modulation spectra, which are the power spectra of a given acoustic parameter sequence.} are hardly perceived by humans. 
Therefore, even if the aliasing occurs on the acoustic parameter sequence, we will not notice it. 
The aliasing issue is a problem specific to the waveform conversion. 
To generate more natural-sounding speech, this aliasing issue remains to be solved.

\subsection{Improved Generator: Addressing Aliasing Issue}

To alleviate the aliasing described in Sec.~\ref{subsec:aliasing}, we have two options. 
One is to explicitly add anti-aliasing processing into the model architecture. 
Following this concept, a linear pooling with Gaussian weights has been proposed~\cite{mairal2014convolutional}. 
This pooling operation is equivalent to the down-sampling after Gaussian filtering. 
We can regard the Gaussian filtering as the approximation of low-pass filtering using the cardinal sine function (a.k.a., sinc function) so that the aliasing will be alleviated well. 
However, there is a fundamental trade-off between the performance and the combination of shift invariance and anti-aliasing.

Another option is to use a dilated convolution~\cite{chen2014semantic}, which is introduced to the deep learning for semantic image segmentation, rather than the classical convolution with strides. 
This is a technique to reduce the number of model parameters and obtain the computational efficiency while maintaining a large receptive field to cater for long-range dependencies. 
Note that the recent neural-network-based vocoder such as WaveNet has also adopted the dilated convolution. 
Toward a high-quality neural post-filter for speech waveform generation, assuming that down/up-sampling modules are not suitable for the speech waveform conversion unlike acoustic parameter conversion, we replace the classical convolution with the dilated convolution in the architecture of WaveCycleGAN.

\subsection{Improved Discriminator: Multiple Domains}
\label{subsec:disciminator}
In the preliminary experiment, although using the dilated convolution instead of the convolution with strides made it possible to alleviate the aliasing, the processed speech somehow became noisy speech, as shown in Fig.~\ref{fig:spectrum} (d). 
Theoretically, the generator should imitate a $\pdf$ of the real data if the training succeeds. 
However, in practice, the gradient of the generator vanishes when the discriminator successfully rejects generated samples with high confidence. 
For this reason, we used the discriminators that have small model parameters, but this insufficient capability of the discriminator might make the decision boundaries non-optimal.

To find the best decision boundaries while avoiding the vanishing gradient problem of the generator, we propose discriminators combining multiple domains such as the waveform domain $D_{Y_\mathrm{wave}}$ and Mel spectrogram domain $D_{Y_\mathrm{msp}}$ as follows:
\begin{align}
  {\mathcal L}_\mathrm{adv} & (G_{X\to Y}, D_Y) = \mathbb{E}_{y \sim P_Y(y)} [\log D_{Y_\mathrm{wave}} (\y)] \nonumber \\
     & + \mathbb{E}_{y \sim P_Y(y)} [\log D_{Y_\mathrm{msp}} (F(\y))] \nonumber \\
     & + \mathbb{E}_{x \sim P_X(x)} [\log (1 - D_{Y_\mathrm{wave}} (G_{X\to Y}(\x)))] \nonumber \\
     & + \mathbb{E}_{x \sim P_X(x)} [\log (1 - D_{Y_\mathrm{msp}} (F(G_{X\to Y}(\x))))]. \label{eq:loss_adv}
\end{align}
\noindent where $F$ indicates a linear mapping function described as a convolution of the Hanning window, followed by a fast Fourier transform (FFT) matrix and Mel-filter bank. 
Unlike the L1 and L2 losses on spectra~\cite{oord2017parallel,takaki2018stft,wang2018neural}, we use the adversarial losses for the multiple domains, so the objective function related to the generator still does not depend directly on $\y$ at all and our approach makes this objective function resistant to the over-smoothing problems~\cite{zen2009statistical} the same as conventional WaveCycleGAN.

\section{Experiments}

\subsection{SPSS using Internal Japanese Dataset}

\subsubsection{Dataset}

We used a Japanese speech dataset consisting of utterances by one professional female narrator. 
To train the models, we used about 6,500 sentences for a baseline system and 400 sentences (speech sections of 1.2 hours) each for WaveCycleGAN and WaveCycleGAN2. 
To evaluate the performance, we used 30 sentences (speech sections of 5.3 minutes). 
The sampling rate of the speech signals was 22.05 kHz.

\subsubsection{Systems}
\label{subsubsec:systems}
We used a deep neural network (DNN)-based SPSS~\cite{zen2013statistical} and WaveCycleGAN~\cite{tanaka2018wavecyclegan} as a baseline system ({\bf SPSS}) and a conventional system ({\bf V1}). 
As a proposed system, {\bf V2} indicates WaveCycleGAN2, which has only the speech waveform domain's discriminators. 
{\bf V2+} indicates WaveCycleGAN2 incorporating the discriminators of the acoustic parameter domains such as the Mel spectrogram ({\bf V2msp}), Mel-frequency cepstrum coefficients ({\bf V2mfcc}), and phase spectrogram ({\bf V2ph}). 
The architecture of the generator was a linear projection (\# of channel, kernel, dilation: 64, 15, 1) followed by a residual block (128, 15, 2), five residual blocks (128, 15, 4), and a linear projection (1, 15, 1). 
We applied the conventional and proposed systems to the speech waveform {\bf SPSS}. 
We used the same learning rate for the first 80k iterations and linearly decayed to 0 over the next 80k iterations. 
The other conditions are the same as in our previous work~\cite{tanaka2018wavecyclegan}.

\subsubsection{Objective Evaluation}
To evaluate the capability of addressing the over-smoothing effect caused by the SPSS, we calculated modulation spectrum differences (MSD) for the Mel cepstral coefficient of natural speech ({\bf Recorded}). 
Although the modulation spectrum is traditionally defined as a value calculated using the Fourier transform of the parameter sequence~\cite{Atlas2003}, this paper defines the modulation spectrum as its logarithmic power spectrum. 
We used 8,192 FFT points.

\begin{figure}[t]
    \centering
    \includegraphics[trim=0.0cm 0.0cm 0.0cm 0.0cm, width=80mm,clip]{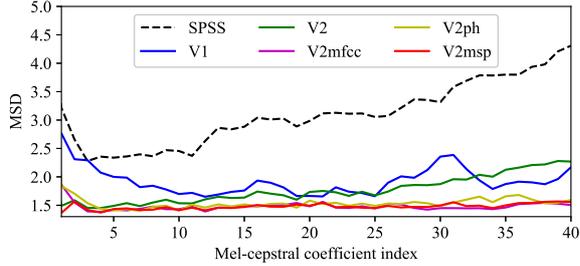}
    \caption{Modulation spectrum differences against for Mel cepstral coefficient of natural speech.}
    \label{fig:msd}
\end{figure}

Figure~\ref{fig:msd} showed that {\bf SPSS} significantly suffered from the over-smoothing effect. 
Although {\bf V1} alleviated its over-smoothing effect, there was still a gap. 
On the other hand, {\bf V2+} restored the modulation spectrum of {\bf Recorded} well. 
Note that as described in Sec.~\ref{subsec:disciminator}, the speech generated by {\bf V2} was more different from natural speech than that generated by the combination methods {\bf V2+}.

\subsubsection{Subjective Evaluation}
\label{subsubsec:spssresult}
We conducted a listening test with a 5-scale mean opinion score regarding naturalness. 
On each system, 200 speech samples (10 participants $\times$ 20 randomly selected speech samples) were evaluated.

\begin{figure}[t]
    \centering
    \includegraphics[trim=0.0cm 0.0cm 0.0cm 0.0cm, width=80mm,clip]{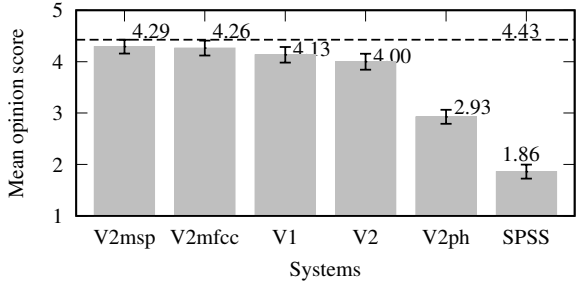}
    \caption{Subjective 5-scale mean opinion scores regarding naturalness, with 95\% confidence intervals. Dashed line indicates results of recorded natural speech.}
    \label{fig:naturalness}
\end{figure}

Figure~\ref{fig:naturalness} showed that {\bf V1} significantly improved the naturalness of the generated speech compared with {\bf SPSS}. 
{\bf V2msp} and {\bf V2mfcc} were closer to natural speech, and there is no statistical difference from natural speech because p values of two-sided Mann-Whitney tests are more than 0.05. 
In contrast, {\bf V2} suffered from noisy speech. 
Note that the score of {\bf V2ph} was significantly degraded because the silence sections somehow became quite noisy. 
These results suggest that it is better to combine the waveform domain discriminator and the amplitude spectrum domain discriminator.

\subsection{Analysis and Synthesis using LJSpeech Dataset~\cite{ljspeech17}}

\subsubsection{Dataset}
We used a public domain English speech dataset~\cite{ljspeech17} containing 13,100 utterances. 
To evaluate the performance, we used 40 sentences disjoint from the training data. 
The sampling rate of the speech signals was 22.05 kHz.

\subsubsection{Systems}
We used WORLD~\cite{morise2016world} and Griffin-Lim~\cite{griffin1984signal} vocoders as the parametric and phase vocoder, respectively. 
For the neural-network-based vocoder, we used open WaveNet~\cite{ryuichiURLwavenet} employing a mixture of logistics distribution~\cite{salimans2017pixelcnn++} and official WaveGlow~\cite{prenger2018waveglow}. 
The audio samples of open WaveNet and official WaveGlow were brought from a public folder~\footnote{\label{foot6} Speech samples can be accessed in a public folder of Google Drive:~\url{http://bit.ly/2JTDetX}} of R. Valle who is a co-author of WaveGlow~\cite{prenger2018waveglow}. 
For the proposed method, we used WaveCycleGAN2 incorporating the Mel spectrum domain discriminator {\bf V2msp}. 
Note that our proposed method worked in both the parallel-data condition {\bf V2msp (paired)} and non-parallel-data condition {\bf V2msp (unpaired)} where the mini-batches of natural speech differed from those of synthesized speech in every iteration.

\subsubsection{Objective and Subjective Evaluations}
To evaluate the capability of WaveCycleGAN2 for an analysis-and-synthesis task, we calculated log spectral distortions (LSD) and conducted a listening test with 5-scale mean opinion scores regarding naturalness. 
On each system, 210 speech samples (14 participants $\times$ 15 randomly selected speech samples) were evaluated subjectively.

The results of LSD, as shown in Tab.~\ref{table:result-lsd-ljspeech-abs}, show that {\bf Griffin-Lim} has the lowest distortion. 
On the other hand, {\bf WORLD} had higher the distortion because of the parametric vocoder. 
In the comparison of {\bf open WaveNet} and {\bf WaveGlow}, {\bf open WaveNet} has larger distortion. 
One possible reason is that {\bf open WaveNet} might generate speech waveforms that have different amplitude spectra from the given acoustic parameter when the previous outputs are captured more strongly than the given acoustic parameters. 
In contrast, {\bf V2msp} has smaller distortion than {\bf WaveGlow}. 
This might be because WaveCycleGAN2 has the advantage of the speech waveform conversion where the input and output domains are closer than those of WaveGlow.

In the results of the listening test, there is no statistical difference in the only two pairs of {\bf WORLD}-{\bf Griffin-Lim} and {\bf open WaveNet}-{\bf V2msp (unpaired)} because p values of two-sided Mann-Whitney tests are more than 0.05. 
Remarkably, {\bf V2msp (paired)} outperformed {\bf open WaveNet} and {\bf WaveGlow}. 
Unlike {\bf WaveGlow}, our proposed method is specified to work on only speech signals whereas {\bf WaveGlow} theoretically works on not only speech signals but also audio signals such as music. 
Moreover, {\bf open WaveNet} is not an official implementation, so this might not be the best result of WaveNet~\cite{oord2016wavenet}. 
However, these results are impressive, and we also had the following feedback from the participants: 
1) {\bf WaveGlow} sometimes had artifacts like {\bf Griffin-Lim}, 
2) {\bf open WaveNet} sometimes had artifacts like the collapsed speech samples reported by Wu et al.~\cite{wu2018collapsed}, 
and 3) {\bf V2msp} sometimes had artifacts caused by the unvoiced/voiced detection error of the WORLD vocoder. 
Note that the tendency of the results compared with {\bf Recorded} differs from the tendency described in Sec.~\ref{subsubsec:spssresult} because the LJSpeech dataset~\cite{ljspeech17} suffers from reverb.

\begin{table}[t]
    \caption{Log spectral distortions (LSD) and subjective 5-scale mean opinion scores regarding naturalness (Naturalness), with 95\% confidence intervals. Mcep indicates Mel cepstral coefficient.}
        \begin{center}
          \begin{tabular}{|l||c|c|} \hline
            System & LSD [dB] & Naturalness \\ \hline \hline
            Recorded~\cite{ljspeech17} & --- & 4.590 $\pm$ 0.082 \\ 
            WORLD~\cite{morise2016world} + mcep & 4.414 $\pm$ 0.022 & 3.124 $\pm$ 0.150 \\
            Griffin-Lim~\cite{griffin1984signal} & 1.546 $\pm$ 0.016 & 3.300 $\pm$ 0.143 \\ 
            open WaveNet (MoL)~\cite{ryuichiURLwavenet} & 4.971 $\pm$ 0.041 & 3.657 $\pm$ 0.162 \\ 
            WaveGlow~\cite{prenger2018waveglow} & 4.540 $\pm$ 0.036 & 3.443 $\pm$ 0.164 \\ \hline
            V2msp (paired) & {\bf 4.318} $\pm$ 0.019 & {\bf 4.023} $\pm$ 0.124 \\
            V2msp (unpaired) & 4.339 $\pm$ 0.020 & 3.833 $\pm$ 0.127 \\ \hline
          \end{tabular}
        \end{center}
    \label{table:result-lsd-ljspeech-abs}
\end{table}

\section{Conclusions}
We proposed a time-domain neural post-filter for speech waveform generation, WaveCycleGAN2. 
Experimental results demonstrated that the proposed method 
1) outperformed the conventional WaveCycleGAN, 
2) is useful for both speech waveforms generated by analysis-and-synthesis and statistical parametric speech synthesis, 
and 3) generated speech waveforms comparable to those of natural speech and speech synthesized by WaveNet~\cite{oord2016wavenet} (open WaveNet~\cite{ryuichiURLwavenet}) and WaveGlow~\cite{prenger2018waveglow}.

\section{Acknowledgements}
This work was supported by a grant from the Japan Society for the Promotion of Science (JSPS KAKENHI 17H01763). 
The authors thank Ryuichi Yamamoto and the authors of WaveGlow.

\vfill\pagebreak
\bibliographystyle{IEEEtran}

\begin{thebibliography}{10}
\providecommand{\url}[1]{#1}
\csname url@samestyle\endcsname
\providecommand{\newblock}{\relax}
\providecommand{\bibinfo}[2]{#2}
\providecommand{\BIBentrySTDinterwordspacing}{\spaceskip=0pt\relax}
\providecommand{\BIBentryALTinterwordstretchfactor}{4}
\providecommand{\BIBentryALTinterwordspacing}{\spaceskip=\fontdimen2\font plus
\BIBentryALTinterwordstretchfactor\fontdimen3\font minus
  \fontdimen4\font\relax}
\providecommand{\BIBforeignlanguage}[2]{{%
\expandafter\ifx\csname l@#1\endcsname\relax
\typeout{** WARNING: IEEEtran.bst: No hyphenation pattern has been}%
\typeout{** loaded for the language `#1'. Using the pattern for}%
\typeout{** the default language instead.}%
\else
\language=\csname l@#1\endcsname
\fi
#2}}
\providecommand{\BIBdecl}{\relax}
\BIBdecl

\bibitem{oord2016wavenet}
A.~v.~d. Oord, S.~Dieleman, H.~Zen, K.~Simonyan, O.~Vinyals, A.~Graves,
  N.~Kalchbrenner, A.~Senior, and K.~Kavukcuoglu, ``{WaveNet}: A generative
  model for raw audio,'' \emph{arXiv preprint arXiv:1609.03499}, 2016.

\bibitem{ryuichiURLwavenet}
R.~Yamamoto, ``{WaveNet} vocoder,'' in
  \emph{\url{https://doi.org/10.5281/zenodo.1472609}}.

\bibitem{prenger2018waveglow}
R.~Prenger, R.~Valle, and B.~Catanzaro, ``{WaveGlow}: A flow-based generative
  network for speech synthesis,'' \emph{arXiv preprint arXiv:1811.00002}, 2018.

\bibitem{kawahara1999restructuring}
H.~Kawahara, I.~Masuda-Katsuse, and A.~De~Cheveigne, ``Restructuring speech
  representations using a pitch-adaptive time-frequency smoothing and an
  instantaneous-frequency-based {F0} extraction: Possible role of a repetitive
  structure in sounds,'' \emph{Speech communication}, vol.~27, no.~3, pp.
  187--207, 1999.

\bibitem{kawahara2001aperiodicity}
H.~Kawahara, J.~Estill, and O.~Fujimura, ``Aperiodicity extraction and control
  using mixed mode excitation and group delay manipulation for a high quality
  speech analysis, modification and synthesis system {STRAIGHT},'' in
  \emph{Second International Workshop on Models and Analysis of Vocal Emissions
  for Biomedical Applications}, 2001.

\bibitem{morise2016world}
M.~Morise, F.~Yokomori, and K.~Ozawa, ``{WORLD}: a vocoder-based high-quality
  speech synthesis system for real-time applications,'' \emph{IEICE
  TRANSACTIONS on Information and Systems}, vol.~99, no.~7, pp. 1877--1884,
  2016.

\bibitem{zen2013statistical}
H.~Zen, A.~Senior, and M.~Schuster, ``Statistical parametric speech synthesis
  using deep neural networks,'' in \emph{Acoustics, Speech and Signal
  Processing (ICASSP), 2013 IEEE International Conference on}.\hskip 1em plus
  0.5em minus 0.4em\relax IEEE, 2013, pp. 7962--7966.

\bibitem{toda2007voice}
T.~Toda, A.~W. Black, and K.~Tokuda, ``Voice conversion based on
  maximum-likelihood estimation of spectral parameter trajectory,'' \emph{IEEE
  Transactions on Audio, Speech, and Language Processing}, vol.~15, no.~8, pp.
  2222--2235, 2007.

\bibitem{zen2009statistical}
H.~Zen, K.~Tokuda, and A.~W. Black, ``Statistical parametric speech
  synthesis,'' \emph{Speech Communication}, vol.~51, no.~11, pp. 1039--1064,
  2009.

\bibitem{tanaka2018wavecyclegan}
K.~Tanaka, T.~Kaneko, N.~Hojo, and H.~Kameoka, ``{WaveCycleGAN}:
  Synthetic-to-natural speech waveform conversion using cycle-consistent
  adversarial networks,'' \emph{arXiv preprint arXiv:1809.10288}, 2018.

\bibitem{kominek2004cmu}
J.~Kominek and A.~W. Black, ``The {CMU} {Arctic} speech databases,'' in
  \emph{Fifth ISCA workshop on speech synthesis}, 2004.

\bibitem{ljspeech17}
K.~Ito, ``The {LJ} speech dataset,''
  \url{https://keithito.com/LJ-Speech-Dataset/}, 2017.

\bibitem{kalchbrenner2018efficient}
N.~Kalchbrenner, E.~Elsen, K.~Simonyan, S.~Noury, N.~Casagrande, E.~Lockhart,
  F.~Stimberg, A.~v.~d. Oord, S.~Dieleman, and K.~Kavukcuoglu, ``Efficient
  neural audio synthesis,'' \emph{arXiv preprint arXiv:1802.08435}, 2018.

\bibitem{wang2018neural}
X.~Wang, S.~Takaki, and J.~Yamagishi, ``Neural source-filter-based waveform
  model for statistical parametric speech synthesis,'' \emph{arXiv preprint
  arXiv:1810.11946}, 2018.

\bibitem{wu2018collapsed}
Y.-C. Wu, K.~Kobayashi, T.~Hayashi, P.~L. Tobing, and T.~Toda, ``Collapsed
  speech segment detection and suppression for wavenet vocoder,'' \emph{arXiv
  preprint arXiv:1804.11055}, 2018.

\bibitem{oord2017parallel}
A.~v.~d. Oord, Y.~Li, I.~Babuschkin, K.~Simonyan, O.~Vinyals, K.~Kavukcuoglu,
  G.~v.~d. Driessche, E.~Lockhart, L.~C. Cobo, F.~Stimberg \emph{et~al.},
  ``Parallel {WaveNet}: Fast high-fidelity speech synthesis,'' \emph{arXiv
  preprint arXiv:1711.10433}, 2017.

\bibitem{ping2018clarinet}
W.~Ping, K.~Peng, and J.~Chen, ``{ClariNet}: Parallel wave generation in
  end-to-end text-to-speech,'' \emph{arXiv preprint arXiv:1807.07281}, 2018.

\bibitem{tobing2019voice}
P.~L. Tobing, Y.-C. Wu, T.~Hayashi, K.~Kobayashi, and T.~Toda, ``Voice
  conversion with cyclic recurrent neural network and fine-tuned {WaveNet}
  vocoder,'' in \emph{ICASSP2019}, 2019.

\bibitem{takamichi2014postfilter}
S.~Takamichi, T.~Toda, G.~Neubig, S.~Sakti, and S.~Nakamura, ``A postfilter to
  modify the modulation spectrum in {HMM}-based speech synthesis,'' in
  \emph{Acoustics, Speech and Signal Processing (ICASSP), 2014 IEEE
  International Conference on}.\hskip 1em plus 0.5em minus 0.4em\relax IEEE,
  2014, pp. 290--294.

\bibitem{kaneko2017generative}
T.~Kaneko, H.~Kameoka, N.~Hojo, Y.~Ijima, K.~Hiramatsu, and K.~Kashino,
  ``Generative adversarial network-based postfilter for statistical parametric
  speech synthesis,'' in \emph{Proc. 2017 IEEE International Conference on
  Acoustics, Speech and Signal Processing (ICASSP2017)}, 2017, pp. 4910--4914.

\bibitem{saito2018statistical}
Y.~Saito, S.~Takamichi, H.~Saruwatari, Y.~Saito, S.~Takamichi, and
  H.~Saruwatari, ``Statistical parametric speech synthesis incorporating
  generative adversarial networks,'' \emph{IEEE/ACM Transactions on Audio,
  Speech and Language Processing (TASLP)}, vol.~26, no.~1, pp. 84--96, 2018.

\bibitem{ma2018a}
S.~Ma, D.~Mcduff, and Y.~Song, ``A generative adversarial network for style
  modeling in a text-to-speech system,'' in \emph{International Conference on
  Learning Representations}, 2019.

\bibitem{kameoka2018stargan}
H.~Kameoka, T.~Kaneko, K.~Tanaka, and N.~Hojo, ``{StarGAN-VC}: Non-parallel
  many-to-many voice conversion with star generative adversarial networks,''
  \emph{arXiv preprint arXiv:1806.02169}, 2018.

\bibitem{he2016dual}
D.~He, Y.~Xia, T.~Qin, L.~Wang, N.~Yu, T.~Liu, and W.-Y. Ma, ``Dual learning
  for machine translation,'' in \emph{Advances in Neural Information Processing
  Systems}, 2016, pp. 820--828.

\bibitem{zhu2017unpaired}
J.-Y. Zhu, T.~Park, P.~Isola, and A.~A. Efros, ``Unpaired image-to-image
  translation using cycle-consistent adversarial networks,'' \emph{arXiv
  preprint arXiv:1703.10593}, 2017.

\bibitem{goodfellow2014generative}
I.~Goodfellow, J.~Pouget-Abadie, M.~Mirza, B.~Xu, D.~Warde-Farley, S.~Ozair,
  A.~Courville, and Y.~Bengio, ``Generative adversarial nets,'' in
  \emph{Advances in neural information processing systems}, 2014, pp.
  2672--2680.

\bibitem{zhou2016learning}
T.~Zhou, P.~Krahenbuhl, M.~Aubry, Q.~Huang, and A.~A. Efros, ``Learning dense
  correspondence via 3{D}-guided cycle consistency,'' in \emph{Proceedings of
  the IEEE Conference on Computer Vision and Pattern Recognition}, 2016, pp.
  117--126.

\bibitem{taigman2016unsupervised}
Y.~Taigman, A.~Polyak, and L.~Wolf, ``Unsupervised cross-domain image
  generation,'' \emph{arXiv preprint arXiv:1611.02200}, 2016.

\bibitem{he2016deep}
K.~He, X.~Zhang, S.~Ren, and J.~Sun, ``Deep residual learning for image
  recognition,'' in \emph{Proceedings of the IEEE conference on computer vision
  and pattern recognition}, 2016, pp. 770--778.

\bibitem{ledig2016photo}
C.~Ledig, L.~Theis, F.~Husz{\'a}r, J.~Caballero, A.~Cunningham, A.~Acosta,
  A.~Aitken, A.~Tejani, J.~Totz, Z.~Wang \emph{et~al.}, ``Photo-realistic
  single image super-resolution using a generative adversarial network,''
  \emph{arXiv preprint}, 2016.

\bibitem{sun2018fishnet}
S.~Sun, J.~Pang, J.~Shi, S.~Yi, and W.~Ouyang, ``{FishNet}: A versatile
  backbone for image, region, and pixel level prediction,'' in \emph{Advances
  in Neural Information Processing Systems}, 2018, pp. 762--772.

\bibitem{ravanelli2018speaker}
M.~Ravanelli and Y.~Bengio, ``Speaker recognition from raw waveform with
  {SincNet},'' \emph{arXiv preprint arXiv:1808.00158}, 2018.

\bibitem{zeiler2014visualizing}
M.~D. Zeiler and R.~Fergus, ``Visualizing and understanding convolutional
  networks,'' in \emph{European conference on computer vision}.\hskip 1em plus
  0.5em minus 0.4em\relax Springer, 2014, pp. 818--833.

\bibitem{gong2018impact}
Y.~Gong and C.~Poellabauer, ``Impact of aliasing on deep {CNN}-based end-to-end
  acoustic models,'' \emph{Proc. Interspeech 2018}, pp. 2698--2702, 2018.

\bibitem{shannon1998communication}
C.~E. Shannon, ``Communication in the presence of noise,'' \emph{Proceedings of
  the IEEE}, vol.~86, no.~2, pp. 447--457, 1998.

\bibitem{takamichi2015naist}
S.~Takamichi, K.~Kobayashi, K.~Tanaka, T.~Toda, and S.~Nakamura, ``The {NAIST}
  text-to-speech system for the blizzard challenge 2015,'' in \emph{Proc.
  Blizzard Challenge workshop}, 2015.

\bibitem{mairal2014convolutional}
J.~Mairal, P.~Koniusz, Z.~Harchaoui, and C.~Schmid, ``Convolutional kernel
  networks,'' in \emph{Advances in neural information processing systems},
  2014, pp. 2627--2635.

\bibitem{chen2014semantic}
L.-C. Chen, G.~Papandreou, I.~Kokkinos, K.~Murphy, and A.~L. Yuille, ``Semantic
  image segmentation with deep convolutional nets and fully connected crfs,''
  \emph{arXiv preprint arXiv:1412.7062}, 2014.

\bibitem{takaki2018stft}
S.~Takaki, T.~Nakashika, X.~Wang, and J.~Yamagishi, ``{STFT} spectral loss for
  training a neural speech waveform model,'' \emph{arXiv preprint
  arXiv:1810.11945}, 2018.

\bibitem{Atlas2003}
L.~Atlas and S.~A. Shamma, ``Joint acoustic and modulation frequency,''
  \emph{EURASIP Journal on Advances in Signal Processing}, vol. 2003, no.~7, p.
  310290, June 2003.

\bibitem{griffin1984signal}
D.~Griffin and J.~Lim, ``Signal estimation from modified short-time {F}ourier
  transform,'' \emph{IEEE Transactions on Acoustics, Speech, and Signal
  Processing}, vol.~32, no.~2, pp. 236--243, 1984.

\bibitem{salimans2017pixelcnn++}
T.~Salimans, A.~Karpathy, X.~Chen, and D.~P. Kingma, ``{PixelCNN++}: Improving
  the pixelcnn with discretized logistic mixture likelihood and other
  modifications,'' \emph{arXiv preprint arXiv:1701.05517}, 2017.

\end{thebibliography}

\end{document}